\documentclass[a4paper,11pt]{article}
\usepackage[margin=2.5cm]{geometry}
\usepackage{amsmath}
\usepackage{mathptmx}
\usepackage{latexsym}
\usepackage{bm}
\usepackage[authoryear,longnamesfirst]{natbib}
\usepackage[normalem]{ulem}
\usepackage{float}
\usepackage{graphicx}
\usepackage{xcolor}
\usepackage{afterpage}
\usepackage[authoryear,longnamesfirst]{natbib}
\usepackage{amsthm}

\def\m#1{\mbox{#1}}                

\newcommand{\E}{\mathrm{E}}         
\newcommand{\Cov}{\mathrm{Cov}}
\newcommand{\tr}{{\mathrm{{\tiny T}}}}
\newcommand{\diag}{{\mathrm diag}}
\newtheorem{lemma}{Lemma}
\begin{document}

\title{A new method of Ecological Inference based on the Cressie--Read power divergence}

\author{
Jose M. Pav\'{\i}a\thanks{GIPEyOP, University of Valencia, Spain.
Email: Jose.M.Pavia@uv.es}
\and
Antonio Forcina\thanks{University of Perugia, Italy}
}

\date{}

\maketitle

\begin{abstract}
Ecological inference aims to estimate the bivariate association structure between a pair of variables from the row and column totals of a set of contingency tables. The paper introduces a new approach based on minimizing the Cressie--Read power divergence, applied to the estimation of vote transitions between two elections. It also outlines a Fisher scoring algorithm and derives the asymptotic distribution of the estimator.
\end{abstract}

\section{Introduction}\label{Int}
Studies on the behavior of voters in two close in time elections are used to throw light on several features of interest to political scientists and the media, like the proportion of voters who: (i) remained faithful to a given party, (ii) moved to abstention or to a protest party, (iii) came back from abstention. These quantities could be measured if the bivariate distribution of the options selected by voters in the two elections could be observed; this is not possible due to the secrecy of the ballots.
Estimates from sample surveys, in addition to being expensive, are affected by non response, memory problem in recalling previous decisions or unwillingness to disclose an apparently inconsistent behavior, see \cite{Russo14}.

As an alternative to surveys, several methods of ecological inference have been developed to estimate the joint distribution of voting choices in two elections from the observed marginal totals. \cite{Goodman} provided a statistical foundation for a simple method based on linear regression, which was widely used at the time in Sociology and Epidemiology. A more modern version of the same approach was introduced by \cite{Rosen}, who parameterized conditional probabilities through their corresponding logits, thereby preventing estimates from falling outside the [0, 1] interval.

Several other methods have been developed in the literature; here, we restrict attention to those most closely related to the approach proposed in this paper. The method of \cite{BrownPayne} is based on maximizing the normal approximation to the Dirichlet-multinomial, which accounts for overdispersion \citep{ForcinaPavia2026}; a version of the same model with a modified variance function was proposed by \cite{FoGnBr}. Another line of work, including the methods discussed in \cite{PaviaRomero}, use linear programming techniques to minimize the sum of the absolute differences between observed and predicted counts.

The collection of methods proposed by \cite{Judge2004}, under the heading of {\it Information Theoretic Approach}, estimates transition probabilities by minimizing a version of the Cressie--Read power divergence between the unknown transition probabilities and a reference conditional distribution for each row. Unless prior information is available, they suggest using a uniform distribution as reference, a choice which ignores the tendency of voters to be faithful, to some extent, to the choice made in the earlier election. A related information-theoretic approach, based on Generalized Maximum Entropy (GME), is implemented in the \texttt{eiIT} \texttt{R} package \citep{Pavia2026eiIT}. 

In this paper, we develop a new method of ecological inference in which transition probabilities are estimated by minimizing a version of the Cressie--Read divergence statistic based on comparing observed and expected numbers of voters in the second election in each local unit.
Basic notations and an outline of the method are introduced in section \ref{sec2}; estimation by a Fisher scoring algorithm is described in section \ref{sec3}, asymptotic properties are derived in section \ref{sec4} and the relation with the methods of \cite{Judge2004} is discussed in the final section.
\section{Notations}
\label{sec2}
Suppose we want to estimate vote transitions within a given district comprising $S$ polling stations. Let $R$ denote the number of voting options in the first election and $C$ the number of options in the second. Let $n_{su+}$ denote the number of voters who selected option $u$ in the first election at polling station $s$, and let $n_{s+v}$ denote the number of voters who selected option $v$ in the second election. Assume that transition probabilities are constant across polling stations, and let $P_{uv}$ denote the probability that a randomly selected voter chooses option $v$ in the second election, conditional on having selected option $u$ in the first election.

To constrain the estimated probabilities to lie in $(0,1)$, we use a multinomial logistic parameterization, taking the last category as the reference:
\[
\beta_{uv} = \log\left(\frac{P_{uv}}{P_{uC}}\right)
\quad \Leftrightarrow \quad
P_{uv}=\frac {\exp(\beta_{uv})} {1+\sum_{h=1}^{C-1}\exp(\beta_{uh})},
\]
for $v=1,\ldots,C-1$, and
\[
P_{uC}=1-\sum_{v=1}^{C-1}P_{uv}.
\]
Let $\boldsymbol{\beta}_u$ denote the vector with elements $\beta_{u1},\ldots,\beta_{u,C-1}$. The vectors $\boldsymbol{\beta}_u$ are concatenated to form the parameter vector $\boldsymbol{\beta}$.
Let $n_{s++}$ denote the total number of voters at polling station $s$, and define  $r_{su}$ = $n_{su+}/n_{s++}$,  $c_{sv}$ = $n_{s+v}/ n_{s++}$ and $\tau_{sv}$ = $\sum_u r_{su} P_{uv}$. Under the assumption that voters make their choices independently of one another, $\tau_{sv}$ is the expected proportion of voters choosing option $v$ in the second election, that is, $\rm{E}(c_{sv})=\tau_{sv}$.

Within the present context, the Cressie--Read divergence \citep{CressieRead} may be written as
\[
CR_{\lambda} = \frac{2}{\lambda(\lambda+1)} \sum_s\sum_v y_{sv} \left[\left(\frac{c_{sv}}{\tau_{sv}}\right)^\lambda-1\right],
\]
where $y_{sv}=n_{s+v}$. When $\lambda=0$, the expression above is defined by continuity as
\[
\lim_{\lambda\rightarrow 0}CR_{\lambda} = 2\sum_s\sum_v y_{sv}\log(c_{sv}/\tau_{sv}),
\]
which is twice a weighted sum of Kullback--Leibler divergences within each local unit.

To use matrix notation, let $\bm p_u$ denote the transpose of the $u$th row of the transition matrix $\bm P$, and $\bm\tau_s$ and $\bm c_s$ denote, respectively, the vectors with elements $\tau_{sv}$ and $c_{sv}$, and note that $\bm\tau_s=\sum_{u=1}^R r_{su}\bm p_u$. In the following we use the convention that, if $\bm x$ is a vector, $\bm x^a$ denotes the vector with elements $x_j^a$.

Because the above measures are used here as objective functions to be minimized, multiplicative constants are irrelevant. It is therefore convenient to define, as it will emerge in section \ref{score}, $D_0(\bm\beta) = \sum_s\sum_v y_{sv}\log(c_{sv}/\tau_{sv})$, and, for $\lambda\ne 0,-1$,
\begin{equation}\label{CR-def}
D_\lambda(\bm\beta) = \frac{1}{\lambda(\lambda+1)} \sum_s \bm y_s^\tr
\left[\left(\diag(\bm\tau_s)^{-1}\bm c_s\right)^{\lambda} -\bm 1_C \right].
\end{equation}

When $\lambda=0$, the criterion reduces to, $D_0(\bm\beta) = \sum_s n_{s++}D_{\mathrm{KL}}(\bm c_s\Vert\bm\tau_s)$, a weighted sum of Kullback--Leibler divergences between the observed and expected proportions at each polling station.
\section{Estimation}
\label{sec3}
Transition probabilities may be estimated by minimizing $D_{\lambda}(\bm\beta)$; this can be achieved by a Fisher scoring algorithm with the observed Hessian replaced by its expected value; their expressions are derived next.
\subsection{Score vector and expected second derivative}\label{score}
The score vector can be computed by the chain rule. Let $\bm G$ denote the $C\times(C-1)$ matrix obtained from an identity matrix of size $C$ by deleting its last column; for any vector of probabilities $\bm p$, define $\Omega(\bm p)$ = $\diag(\bm p)-\bm p\bm p^\tr$.
The score corresponding to the $u$th row of the transition matrix is
\[
\bm s_u = \frac{\partial D_\lambda}{\partial\bm \beta_u} =
\frac{\partial\bm p_u^\tr} {\partial\bm \beta_u}\sum_s\frac{\partial\bm \tau_s^\tr} {\partial\bm p_u} \frac{\partial D_\lambda} {\partial\bm \tau_s}
= \bm{G}^\tr\bm{\Omega}(\bm{p}_u) \bm d_u, \,\,\, \m{ with }\bm d_u=\sum_s r_{su}\frac{\partial D_\lambda} {\partial\bm \tau_s}
\]

The vector $\bm d_u$ has size $C$, with elements given by
\[
\bm d_{u} =
\begin{cases}
- \displaystyle\sum_s r_{su} \, \diag(\bm\tau_s)^{-1}\bm y_s  & \lambda=0,\\
- \displaystyle \frac{1}{\lambda+1}\sum_s n_{su+} \, \diag(\bm\tau_s)^{-(\lambda+1)}\bm c_s^{\lambda+1}, & \lambda\ne 0,-1.
\end{cases}
\]
The whole score vector $\bm s$, obtained by stacking the $\bm s_{u}$ vectors one below the other, can be written as
$\bm s$ = $\bm A^\tr \bm d$, where $\bm A$ is a block diagonal matrix with  blocks $\bm\Omega(\bm p_u) \, \bm G$ and the vector $\bm d$ is obtained by stacking the $\bm d_u$ vectors one below the other.

When $\lambda=0$, it is easily verified that $\rm{E}(\bm d_u)$ = $-n_{+u+} \bm 1_C$.  When $\lambda\ne 0$ and the $n_{s++}$ are sufficiently large, in a neighbourhood of $\bm\tau_s$, $\bm d_u$ can be replaced by a first order expansion which gives

\begin{equation} \label{deltaM}
\bm d_u \cong -\sum_s n_{su+} \left[ \frac{1}{\lambda+1}\bm 1_C + \diag(\bm\tau_s)^{-1} (\bm c_s-\bm\tau_s) \right].
\end{equation}

\begin{lemma}
If $\bm p$ is a vector of probabilities summing to one, whose elements are functions of a parameter vector
$\bm \theta$, then
\[
\begin{aligned}
\text{(a)}\quad &
\frac{\partial\bm p^\tr}{\partial\bm{\theta}}\bm 1_C = \bm 0,\\[6pt]
\text{(b)}\quad &
\frac{\partial}{\partial\bm\theta} \left( \frac{\partial\bm p^\tr}{\partial\bm\theta}\bm 1_C \right)
=\bm 0.
\end{aligned}
\]
\end{lemma}
{\sc Proof}. Both identities follow by noting that $\bm p^\tr \bm 1_C = 1$ is a constant with respect to $\bm \theta$, and hence all
its derivatives are zero.$\Box$

Point (a) implies that the expected value of the score vector is zero when $\lambda=0$, and approximately zero otherwise. As we shall see, this result simplifies the computation of the covariance matrix of the score. Point (b) allows us to simplify the computation of the
expected second derivative. Since $\rm{E}(\bm d_u)$ is approximately proportional to $\bm 1_C$ (a condition that holds exactly when $\lambda=0$), the following component of the expected value of the Hessian is approximately zero:
\[
\bm G^\tr \frac{\partial \bm\Omega(\bm p_u)} {\partial\bm\beta_u^\tr} \rm{E}(\bm d_u)\cong\frac{\bm G^\tr \partial\bm\Omega(\bm p_u)\bm 1_C}
{\partial\bm\beta_u^\tr} = \bm 0.
\]

Let $\bm \Delta$ be a matrix consisting of $R\times R$ blocks of the form $\rm{E}(\partial d_u/\partial\bm p_w^\tr)$.
Then the expected Hessian of $D_{\lambda}(\bm \beta)$ is
\[
\bm H = \mathrm{E}\left[\frac{\partial^2 D_{\lambda}(\bm\beta)}{\partial\bm\beta \partial\bm\beta^\tr} \right]
\approx \bm A^\tr \bm\Delta \bm A,
\quad \text{where} \,\,\,\, \bm\Delta_{uw} = \mathrm{E}\left(\frac{\partial\bm d_u}{\partial\bm p_w^\tr}\right) \approx \sum_s \frac{n_{su+}n_{sw+}}{n_{s++}} \diag(\bm\tau_s)^{-1}.
\]
The expressions are exact when $\lambda=0$.
\subsection{An estimation algorithm}
Let $\bm s^{(0)}$ and $\bm H^{(0)}$ denote, respectively, the score vector and the expected Hessian computed at a starting value $\bm\beta^{(0)}$; if $\bm\beta^{(0)}$ is sufficiently close to $\hat{\bm\beta}$, a first-order expansion around $\bm\beta^{(0)}$ gives
\[
\bm 0 = \bm s(\hat{\bm\beta}) \approx \bm s^{(0)} + \bm H^{(0)}(\hat{\bm\beta}-\bm\beta^{(0)}),
\]
which leads to the updating equation $\hat{\bm\beta}=\bm\beta^{(0)} - (\bm H^{(0)})^{-1}\bm s^{(0)}$. In practice, a step-size adjustment may be necessary to ensure convergence.
\section{Asymptotic properties}
\label{sec4}
We consider an asymptotic setting in which the number of polling stations $S$, as well as the numbers of voting options $R$ and $C$, remain fixed, while the total number of voters $n=\sum_s n_{s++}$ tends to infinity. We assume that, for each polling station $s$, $n_{s++}/n\rightarrow\eta_s>0$, and that the row proportions $r_{su}$ converge to constants $\rho_{su}\in(0,1)$. Notice that $\sum_s\eta_s=1$ and $\sum_u\rho_{su}=1$. The transition probabilities are assumed to be strictly positive and constant across polling stations; let $\bm\beta_0$ denote the true value of the parameter vector $\bm\beta$. Finally, we assume that voters make their choices independently of one another, both within and across polling stations.
\subsection{Distribution of the marginal counts}
Let $\bm y_{su}$ denote the vector with elements $n_{suv}$, then, the vector of observed marginal counts may be written as $\bm y_s=\sum_{u=1}^R\bm y_{su}$. Under the assumed data-generating process, $\bm y_{su}\sim \text{Multinomial}(n_{su+},\bm p_u)$, with the vectors $\bm y_{su}$ independent across origin groups and polling stations. Consequently,
\[
\E(\bm y_s)=n_{s++}\bm\tau_s,
\qquad
\Cov(\bm y_s) = \sum_{u=1}^R n_{su+}\bm\Omega(\bm p_u).
\]
Since $S$ and $R$ are fixed and $n_{s++}/n\to\eta_s>0$, the multivariate central limit theorem implies that
\[
\frac{\bm y_s-n_{s++}\bm\tau_s}{\sqrt n} \xrightarrow{d}N(\bm 0,\bm V_s),
\qquad
\bm V_s=\eta_s\sum_{u=1}^R \rho_{su}\bm\Omega(\bm p_u).
\]
\subsection{Asymptotic distribution of the estimator}
When $\lambda=0$, the vector $\bm d_u$ can be written as $\bm d_u$  = $-\sum_{s=1}^S r_{su} \diag(\bm\tau_s)^{-1}\bm y_s$,
therefore, using independence across polling stations, the covariance between $\bm d_u$ and $\bm d_w$ is
\[
\Cov(\bm d_u,\bm d_w) = \sum_{s=1}^S r_{su}r_{sw} \, \diag(\bm\tau_s)^{-1}
\left[ \sum_{h=1}^R n_{sh+} \bm\Omega(\bm p_h) \right] \diag(\bm\tau_s)^{-1}.
\]
Let $\bm\Sigma_d$ = $\lim_{n\to\infty}\mathrm{Cov}(\bm d)/n$, assuming that this limit exists.
Since $\bm d$ is linear in the vectors of marginal counts $\bm y_s$, the multivariate central limit theorem yields
\[
\frac{\bm d-\E(\bm d)}{\sqrt n} \xrightarrow{d}N(\bm 0,\bm\Sigma_d).
\]

Since $\E(\bm d_u)=-n_{+u+}\bm 1_C$ and $\bm\Omega(\bm p_u)\bm 1_C=\bm 0$, it follows that $\E(\bm s_u)=\bm 0$ and hence
$\E(\bm s)=\bm 0$. Let $\bm A_0$ = $\bm A(\bm\beta_0)$ denote the matrix $\bm A$ evaluated at the true parameter value
$\bm\beta_0$, and let $\bm\Sigma_s$  = $\bm A_0^\tr\bm\Sigma_d\bm A_0$. It follows that
\begin{equation}\label{Ascore}
\frac{\bm s(\bm\beta_0)}{\sqrt n} \xrightarrow{d}N(\bm 0,\bm\Sigma_s).
\end{equation}

When $\lambda\ne0,-1$, though $\bm d_u$ is no longer linear in $\bm y_s$, the first order expansion in (\ref{deltaM}) indicates that the first term inside the brackets is proportional to $\bm 1_C$ and therefore vanishes when premultiplied by $\bm A_0^\tr$. Because the second term is the same as in the case of $\lambda=0$, under the usual regularity conditions, (\ref{Ascore}) may be taken to hold in general.

Assume that the usual regularity conditions for asymptotic estimation hold (see, e.g., Section 12.3 of \cite{Wooldrige}), and let $\bm F$ denote the limit of the expected Hessian divided by $n$. Then
\begin{equation}\label{Abeta}
\sqrt n(\hat{\bm\beta}-\bm\beta_0) \xrightarrow{d} N\left(\bm 0, \bm F^{-1}\bm\Sigma_s\bm F^{-1} \right).
\end{equation}

In short, under the stated assumptions, the asymptotic covariance matrix has the usual sandwich form and is the same for all values of $\lambda$. Although the choice of $\lambda$ may affect the estimates in finite samples, it does not affect their asymptotic distribution. In particular, all the criteria in the Cressie--Read family considered here lead to the same first-order asymptotic covariance matrix.
\section{The \cite{Judge2004} approach to ecological inference}
\cite{Judge2004} were the first to consider the Cressie--Read family of divergence measures in the context of ecological inference; their approach, however, differs substantially from the one developed in this paper. In their formulation, the unknown parameters are the transition probabilities $P_{suv}$, that is, the probability of choosing option $v$ in the second election, conditional on having chosen option $u$ in the first election and belonging to local unit $s$, and they propose estimating these probabilities by minimizing a Cressie--Read divergence between the unknown conditional probabilities and a reference distribution.

However, the Cressie--Read power divergence family was originally developed to measure discrepancies between observed and expected frequencies under a specified statistical model \citep{CressieRead}. 
\cite{Judge2004} replace observed counts by unknown conditional probabilities and expected counts with reference probabilities, which are said to be based on ``prior or additional information''. Thus, rather than seeking estimates that make the model-based expected counts as close as possible to the observed ones, their approach seeks transition probabilities that are as close as possible to the reference distribution, subject to the constraints imposed by the observed marginal counts.

On p. 9, they restrict attention to uniform reference probabilities, a choice that could be motivated by the maximum-entropy principle in the absence of additional information. Thus, once a value has been assigned to the $\lambda$ parameter, their approach minimizes the divergence between the candidate transition probabilities and a uniform reference distribution, subject to the constraints that the probabilities sum to one across each row and satisfy the accounting equations $\sum_u n_{su+}P_{suv}=n_{s+v}$ for $v=1,\dots,C$. Within this framework, the method can be applied independently to each local unit. Information sharing across local units would require an additional assumption, namely that $P_{suv}=P_{uv}$ for all $s$.

Let us consider, for simplicity, the case where the method is applied separately to each local unit. Among all tables of conditional probabilities satisfying the accounting equations, it selects the one closest, according to the chosen divergence measure, to a table of uniform transition probabilities. The resulting estimates therefore depend not only on the observed marginal counts but also on the
choice of the uniform reference distribution.

Moreover, the resulting estimates will not, in general, be consistent. Although a larger number of observations makes the estimates of the marginal distributions more accurate, the accounting equations are not sufficient to identify the underlying transition table. Thus, the estimated table will generally converge to the solution selected by the divergence criterion rather than to the true transition table. Consistency would require additional conditions on the structure of the true transition probabilities.

Our approach differs from this strategy. Rather than estimating a separate transition table for each local unit by selecting the one closest to a uniform reference distribution, we assume a common transition matrix across all local units and estimate it by minimizing the
Cressie--Read divergence between the observed marginal counts and the corresponding model-based expected counts. Thus, information from all local units contributes to the estimation of a common set of transition probabilities.

\bibliographystyle{apalike}
\bibliography{CRead}

\end{document}